\documentclass[a4paper]{article}
\usepackage[latin1]{inputenc} 
\usepackage{color}
\usepackage{amsthm}
\usepackage{amsmath}
\usepackage{graphicx}
\usepackage{amssymb}
\usepackage{bbm}
\usepackage{esint}
\usepackage{subfigure}
\usepackage{color}
\usepackage[T1]{fontenc}

\usepackage{anysize}
\marginsize{2cm}{2cm}{2cm}{2cm}

\def\beq{\begin{equation}}
\def\eeq{\end{equation}}
\def\bea{\begin{eqnarray}}
\def\eea{\end{eqnarray}}
\def\nn{\nonumber}

\makeatletter

\theoremstyle{plain}

  \theoremstyle{remark}

\title{Deformed quantum double realization of the toric code and beyond}

\author{Pramod Padmanabhan\footnote{pramod23phys@gmail.com}, Juan Pablo Ibieta Jimenez \footnote{pibieta@if.usp.br}, Miguel Jorge Bernab\'{e} Ferreira \footnote{migueljb@if.usp.br}, \\ Paulo Teotonio-Sobrinho \footnote{teotonio@if.usp.br}  \\ \\
\small{Departmento de F\'{i}sica Matem\'{a}tica Universidade de S\~{a}o Paulo - USP,} \\
\small{ CEP 05508-090 Cidade Universit\'aria, S\~{a}o Paulo - Brasil}}

\date{\today}
\begin{document}

\maketitle

\begin{abstract}

Two dimensional lattice models such as the quantum double models, which includes the toric code, can be constructed from transfer matrices of lattice gauge theories with discrete gauge groups. These transfer matrices are built out of local operators acting on links, vertices and plaquettes and are parametrized by the center of the gauge group algebra and its dual. For general choices of these parameters the transfer matrix contains operators acting on links which can also be thought of as perturbations to the quantum double model driving it out of its topological phase towards a paramagnetic phase. These perturbations can be thought of as magnetic fields added to the system which destroy the exact solvability of the quantum double model. We modify these transfer matrices with perturbations and extract exactly solvable models which remain in a quantum phase, thus nullifying the effect of the perturbation. The algebra of the modified vertex and plaquette operators now obey a deformed version of the quantum double algebra. The Abelian cases are shown to be in the quantum double phase whereas the non-Abelian phases are shown to be in a modified phase of the corresponding quantum double phase. This is shown by working with the groups $\mathbb{Z}_n$ and $S_3$ for the Abelian and non-Abelian cases respectively. The quantum phases are determined by studying the excitations of these systems. The fusion rules and the statistics of these anyons indicate the quantum phases of these models. The implementation of these models can possibly improve the use of quantum double models for fault tolerant quantum computation. We then construct theories which arise from transfer matrices that are not the transfer matrices of lattice gauge theories. In particular we show that for the $\mathbb{Z}_2$ case this contains the double semion model. More generally for other discrete groups these transfer matrices contain the twisted quantum double models. These transfer matrices can be thought of as being obtained by introducing extra parameters into the transfer matrix of lattice gauge theories. These parameters are central elements belonging to the tensor products of the algebra and its dual and are associated to vertices and volumes of the three dimensional lattice. As in the case of the lattice gauge theories we construct the operators creating the excitations in this case and study their braiding and fusion properties.  

\end{abstract}

\section{Introduction}

The concept of topological order~\cite{WenBook} was initiated in the 80's with the discovery of the fractional quantum Hall effect and high temperature superconductivity~\cite{DSP, PG, PWA}. Since then it has also been seen in short range resonating valence bond states~\cite{7,8,9,10} and in quantum spin liquids~\cite{11,12,13,5,14,15,16,17}. Once it's usefulness was realized in solid state quantum computation~\cite{NayakReview} several exactly solvable models have been constructed achieving this. The classic example emerged when Kitaev wrote down the toric code Hamiltonian in two dimensions~\cite{KitToric}. These systems are examples of lattice models which host anyons~\cite{WilczekBook} as part of their low energy excitations. They can also be thought of as particular phases of the $\mathbb{Z}_2$ lattice gauge theory which host these deconfined excitations with anyonic statistics. They posses ground states with degeneracies which are topological invariants. This degeneracy is stable up to the addition of weak perturbations. This feature makes this model a probable candidate for fault tolerant quantum computation~\cite{KitToric}. These models have been extended to other discrete groups and involutary Hopf algebras as well~\cite{KitToric, AguQDM}. In these cases they can be thought of as arising from lattice gauge theories based on these discrete groups or involutary Hopf algebras~\cite{QDMST}. Earlier works showing the existence of anyons in two dimensional discrete gauge theories can be found in~\cite{Bais, Bais1}. Discrete gauge theories emerge in these models when a continuous gauge group breaks down to one via spontaneous symmetry breaking~\cite{22, 23, 24, 25, 26, 27, 28}. 

These systems are usually perturbed by adding qudit terms to act on the edges of the lattice which carry the gauge degrees of freedom. They drive the system out of the topological phase~\cite{Castel1, CC1, Bravyi1, Bravyi2, Tsomokos, CC2, Vidal}. The resulting models are rendered unsolvable analytically and are thus subject to study using numerical methods. However by considering restricted plaquette and vertex operators they can be made solvable. Such studies were carried out in~\cite{Del1}. This resulted in condensed phases of the quantum double model. These works were crucial in understanding the stability of these topological phase represented by the quantum double model. It is thus an important problem to find exactly solvable models which remain in topological phases in the presence of these perturbations.

In this spirit we introduce exactly solvable models which are constructed by looking at possible Hamiltonians that can be generated using the transfer matrices of such systems. We show exactly solvable models which include the single qudit perturbations. The cases of Abelian and non-Abelian groups are studied separately. It is shown that in the Abelian case the system remains in the topological phase corresponding to the quantum double model. The situation turns out different for the non-Abelian cases. We find that the model remains in a topological phase but it is in a modified version with respect to the corresponding topological phase of the quantum double model. These are seen by studying the examples of the group algebras of $\mathbb{Z}_n$ and $S_3$ denoted by $\mathbb{C}(\mathbb{Z}_n)$ and $\mathbb{C}(S_3)$ respectively. 

We then go beyond the transfer matrices of lattice gauge theories by introducing more parameters in the transfer matrix of lattice gauge theories to find other topological phases. This transfer matrix is made up of two-qudit operators apart from the usual operators making up the transfer matrix of lattice gauge theories. We work with the $\mathbb{C}(\mathbb{Z}_2)$ case to show how one can obtain the double semion phase from such considerations. For more general groups these transfer matrices contain the twisted quantum double models as defined in~\cite{YSWu}. Such models were also defined in~\cite{Ran} while considering trivial global symmetry groups. 

The paper is organized as follows. Section 2 describes the transfer matrices of two dimensional lattice gauge theories. The local operators used to build these transfer matrices and the parameters used to study phase transitions between different phases is explained here. Hamiltonians of two dimensional lattice models are obtained from these transfer matrices by taking their logarithms. This can be done in different ways resulting in several models. These include exactly solvable models and models which cannot be solved analytically. The models of interest are presented in section 3. Their excitations along with their braiding statistics and fusion rules are studied in section 4. Section 5 shows the construction of the double Semion model from the transfer matrix picture. Section 6 makes up our concluding remarks. 

\section{The Transfer Matrix}

The system is defined on a two dimensional lattice $\Sigma$ with gauge degrees of freedom located on the links of this lattice. The gauge degrees of freedom belong to the group algebra of a gauge group $G$, denoted by $\mathbb{C}(G)$,  with basis elements $\{\phi_g | g\in G\}$ and multiplication rule given by $\phi_g\phi_h = \phi_{gh}$. Thus the states on the links are given by linear combinations of $|g\rangle$. In the case of $\mathbb{C}(\mathbb{Z}_2)$ this is nothing but the familiar two state system given by $|1\rangle$ and $|-1\rangle$. 

The system is evolved in time resulting in a three dimensional manifold $M$ which is more precisely given by $M=\Sigma\times [0,1]$ where $[0,1]$ is the unit time interval. $M$ is also discretized with elements of the algebra $\mathbb{C}(G)$ living on the links. This can be used to construct partition functions for this system by associating weights to links and faces of the closed triangulated 3-manifold $M$ in the spirit of state-sum models~\cite{state}. However when $M$ is not closed the tensors associated to the weights have non contracted indices at the ends of the unit time interval giving us transfer matrices. Such a procedure was used to construct transfer matrices of lattice gauge theories in~\cite{QDMST}.  

In~\cite{QDMST} the partition function, and hence the transfer matrix, was built out of structure constants of the algebra $\mathcal{A}$, which in this case is given by $\mathbb{C}(G)$, as the weights associated to links and plaquettes. Apart from these structure constants the partition function is also parametrized by a pair of elements $z_S, z_T$ belonging to the center of the algebra $\mathcal{A}$ and $z_S^*, z_T^*$ belonging to the center of the dual algebra $\mathcal{A}^*$. $S$ and $T$ denote space and time directions. Thus the transfer matrix in our case is given by $ U(\mathcal{A}, z_S, z_T, z_S^*, z_T^*)$. We do not go into the details of constructing this transfer matrix in this paper. The details can be found in~\cite{QDMST}. For the purposes of this paper we will start with the most general transfer matrix for $\mathcal{A}= \mathcal{C}(G)$ with particular choices of $G$ and show how we can construct various models of physical interest from them.

\section*{$\mathbb{C}(G)$:}

For a general group algebra based on a group $G$ the transfer matrix is written as 
 \bea\label{tg} U\left(\mathbb{C}(G), z_S, z_T, z_S^*, z_T^*\right) & = &  \prod_p \left(\sum_{C\in [G]} \beta_CB_p^C\right) \nn \\ 
 & \times & \prod_j\left(\sum_{g\in G} b_gT_j^g\right) \prod_i \left(\sum_{R\in \textrm{IRR's of $G$}} a_RX_i^R\right)  \prod_v A_v \eea
 where $p$, $j$, $i$ and $v$ denotes plaquettes, links and vertices respectively. The operators in Eq.(\ref{tg}) are given by 
 \bea \label{operatorsG} A_v & = & \sum_{g\in G}\alpha_g~L_{i_1}^g\otimes R_{i_2}^{g^{-1}}\otimes L_{i_3}^g\otimes R_{i_4}^{g^{-1}} \\
 \label{operatorsBG} B_p^C & = & \sum_{\{g_i\}}\delta\left(g_1 g_2 g_3 g_4,C \right) T_{j_1}^{g_1}\otimes T_{j_2}^{g_2}\otimes T_{j_3}^{g_3^{-1}}\otimes T_{j_4}^{g_4^{-1}} \\
\label{operatorsXG} X_i^R & = & \frac{1}{|G_R|}\sum_{g\in G}\chi_R(g) L_i^g  \eea
 where $\chi_R(g)$ is the character of the group element $g$ in the IRR $R$ , $|G_R|$ is the number of elements with non-zero character in the IRR $R$, $[G]$ is the set of conjugacy classes in $G$, and the operators $T_i^{g^{\pm1}}, L_i^g, R_i^{g^{-1}}$ act on the states $|k\rangle$ on the links as follows
 \bea T_i^g|k\rangle & = & \delta_{g,k}|k\rangle ;~ T_i^{g^{-1}}|k\rangle = \delta_{g^{-1}, k}|k\rangle, \\
 L_i^g|k\rangle & = & |gk\rangle ;~ R_i^{g^{-1}}|k\rangle = |kg^{-1}\rangle.\eea 
 
 The operator in Eq.(\ref{operatorsXG}) can also be defined using the $R_i^{g^{-1}}$ operators instead of the $L_i^g$ operators. However this does not matter as it can be shown that they are the same. This is because the elements of a given conjugacy class have the same coefficients. So the orientation of the lattice does not matter for the definition of these operators. However in the operator $\left(\sum_{g\in G}b_g T_j^g\right)$ the coefficients of the elements in a given conjugacy class are not the same and hence the orientation of the lattice does matter in this case. The convention is that if the orientation of the link matches with the orientation of the plaquette then we use $T^g_j$ otherwise we use $T^{g^{-1}}_j$.   
 
 The parameters on the two sides of Eq.(\ref{tg}) are related by the following 
 \bea z_S & = & \sum_C\beta_C\sum_{g\in C}\phi_g \\ 
 z_T & = & \sum_R a_R \sum_{g\in G}\chi_R(g) \phi_g \\ 
 z_S^* & = & \sum_{g\in G}b_g\psi_g \\
 z_T^* & = &  \sum_{g\in G} \alpha_g\psi_g \eea
 where $\{\psi_g\}$ is the basis of the algebra dual to $\mathbb{C}(G)$ that is it is dual to the basis of the algebra given by $\{\phi_g\}$. 
 
 These are the most general form for the central elements of a group algebra and its dual. In this
paper we are not using $z_S,z_T,z^*_S,z^*_T$ directly. Instead,
we describe the models in terms of $\beta_C,a_R,b_g$ and $\alpha_g$.
The parametrization by algebra and dual algebra elements is important
in the context of~\cite{QDMST} but it will not play a major role here.
 
\begin{figure}[h!]
  \centering
  \includegraphics[scale=1.2]{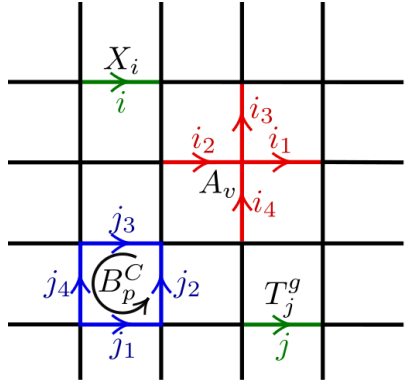}
  \caption{The action of the vertex operator $A_v$ and plaquette operator $B_p^C$ are shown along with the orientations of the lattice.}\label{lattop}
 \end{figure}

 The action of the operators in Eq.(\ref{operatorsG}-\ref{operatorsXG}) is shown in fig.(\ref{lattop}). 
 
 Note from the form of the terms in the transfer matrix in Eq.(\ref{tg}) that the operators on the plaquettes $p$, are the same as the plaquette operators of the quantum double model~\cite{KitToric, AguQDM}. However the operators on the vertices $v$ are not the same as the star operators of the quantum double models. In order to obtain precisely the star operators of the quantum double case we need to choose $z_T^*$ to be
 \beq z_T^* = \sum_{R\in \textrm{IRR's of}~G}\alpha_R\sum_{g\in G}\chi_R(g)\psi_g.\eeq
 This results in the following vertex operator 
 \beq A_v = \sum_{R\in \textrm{IRR's of}~G} \alpha_R A_v^R \eeq
 with 
 \beq A_v^R = \frac{1}{|G_R|}\sum_{g\in G}\chi_R(g)~L_{i_1}^g\otimes R_{i_2}^{g^{-1}}\otimes L_{i_3}^g\otimes R_{i_4}^{g^-{1}}. \eeq 
 
 The operators $A_v^R$ commute with $B_p^C$. When the IRR $R$ is the identity representation we obtain the star operator projecting to the vacuum sector. This is the operator which appears in the Hamiltonian of the quantum double model~\cite{KitToric, AguQDM}. Note that these subtleties do not matter for the Abelian models. They only make a difference in the non-Abelian instances.

 The operators acting on the links $i$, $j$ in Eq.(\ref{tg}) can be thought of as magnetic field terms. Such terms were considered in~\cite{Del1, Del2} to induce condensations of quasiparticle excitations leading to phase transitions while remaining in the exactly solvable regime. 
 
 We now write down the transfer matrices for the examples we consider in this paper.  

\section*{$\mathbb{C}(\mathbb{Z}_2)$:}

We now look at the simple example of $G=\mathbb{Z}_2$ to illustrate the form of the transfer matrix defined in Eq.(\ref{tg}). 
\bea\label{tz2} U(\mathbb{C}(\mathbb{Z}_2), z_S, z_T, z_S^*, z_T^*) & = & \prod_p \left(\beta_1 B_p^1 + \beta_{-1} B_p^{-1}\right)\prod_j \left(b_1 T_j^1 + b_{-1} T_j^{-1}\right) \nn \\ & \times & \prod_i \left(a_1 X_i^1 + a_{-1} X_i^{-1}\right) \prod_v\left(\alpha_1 A_v + \alpha_{-1} A_v^{-1}\right)\eea    
where $v$, $p$, $i$ and $j$ label vertices, plaquettes and links respectively. The operators $A_v^{\pm1}$, $B_p^{\pm 1}$, $X_i^{\pm 1}$ and $T_j^{\pm 1}$ are all projectors and are given by
\bea A_v^{\pm 1} & = & \frac{\mathbbm{1}\otimes\mathbbm{1}\otimes\mathbbm{1}\otimes\mathbbm{1} \pm \sigma^x_{i_1}\otimes\sigma^x_{i_2}\otimes\sigma^x_{i_3}\otimes\sigma^x_{i_4}}{2} \\
        B_p^{\pm 1} & = & \frac{\mathbbm{1}\otimes\mathbbm{1}\otimes\mathbbm{1}\otimes\mathbbm{1} \pm \sigma^z_{j_1}\otimes\sigma^z_{j_2}\otimes\sigma^z_{j_3}\otimes\sigma^z_{j_4}}{2} \\
        X_i^{\pm 1} & = & \frac{\mathbbm{1} \pm \sigma^x_i}{2} \\
        T_j^{\pm 1} & = & \frac{\mathbbm{1} \pm \sigma^z_j}{2}. \eea
 These are the operators appearing in the toric code~\cite{KitToric, AguQDM} system and their action is shown in the figure(\ref{lattop}). 
 
 The parameters on both sides of Eq.(\ref{tz2}) are related in the following way
 \bea z_S & = & \beta_1 \phi_1 + \beta_{-1}\phi_{-1} \\ 
         z_T & = & \left(\frac{a_1+a_{-1}}{2}\right) \phi_1 + \left(\frac{a_1 -a_{-1}}{2}\right) \phi_{-1} \\ 
         z_S^* & = & b_1 \psi_1 + b_{-1} \psi_{-1} \\
         z_T^* & = & \left(\frac{\alpha_1+\alpha_{-1}}{2}\right) \psi_1 + \left(\frac{\alpha_1-\alpha_{-1}}{2}\right) \psi_{-1} \eea
 where $\phi_{\pm 1}$ and $\psi_{\pm 1}$ are the basis elements of $\mathbb{C}(\mathbb{Z}_2)$ and its dual respectively. 
 
 \section*{$\mathbb{C}(\mathbb{Z}_n)$:}
 
 The elements of the group $\mathbb{Z}_n$ are denoted by $\omega^l$ with $l\in\left(0,\cdots, n-1\right)$, the elements of the group algebra $\mathbb{C}(\mathbb{Z}_n)$ and its dual are denoted by $\phi_{\omega^l}$ and $\psi_{\omega^l}$ respectively. 
 
  In this case the transfer matrix is written as
 \bea\label{tzn} U(\mathbb{C}(\mathbb{Z}_n), z_S, z_T, z_S^*, z_T^*) & = & \prod_p\left(\sum_{k=0}^{n-1}\beta_{\omega^k} B_p^{\omega^k}\right)\prod_j\left(\sum_{k=0}^{n-1}b_{\omega^k}T_j^{\omega^k}\right) \nn \\ 
 & \times & \prod_i\left(\sum_{k=0}^{n-1}a_{\omega^k}X_i^{\omega^k}\right)\prod_v\left(\sum_{k=0}^{n-1}\alpha_{\omega^k} A_v^{\omega^k}\right) \eea
 where the index $\omega^k$ labels the IRR of $\mathbb{Z}_n$. The operators $A_v^{\omega^k}$, $B_p^{\omega^k}$, $X_i^{\omega^k}$ and $T_j^{\omega^k}$ are given by
 \bea A_v^{\omega^k} & = & \frac{1}{n}\sum_{l=0}^{n-1}\chi_{\omega^k}(\omega^l)X_{i_1}^l\otimes X_{i_2}^{-l}\otimes X_{i_3}^l\otimes X_{i_4}^{-l} \\
 B_p^{\omega^k} & = & \frac{1}{n}\sum_{l=0}^{n-1}\chi_{\omega^k}(\omega^l)Z_{j_1}^l\otimes Z_{j_2}^l\otimes Z_{j_3}^{-l}\otimes Z_{j_4}^{-l} \\
 X_i^{\omega^k} & = & \frac{1}{n}\sum_{l=0}^{n-1}\chi_{\omega^k}(\omega^l)X_i^l \\
 T_j^{\omega^k} & = & \frac{1}{n}\sum_{l=0}^{n-1} \chi_{\omega^k}(\omega^l)Z_j^l \eea
 where $X_i^l$ and $Z_j^l$ are single qudit operators generating $\mathbb{Z}_n$ and are defined by
 \bea\label{genxn} X_i |\omega^k\rangle & = & |\omega^{k+1}\rangle \\ 
\label{genzn} Z_j|\omega^k\rangle & = & \omega^k|\omega^k\rangle.\eea
 
 $\chi_{\omega^k}(\omega^l)$ is the character of the element $\omega^l$ in the IRR labeled by $\omega^k$.
 
 The parameters on the two sides of Eq.(\ref{tzn}) are related in the following way
 \bea z_S & = & \sum_{k=0}^{n-1}\beta_{\omega^k}\phi_{\omega^k} \\ 
 z_T & = & \sum_{k=0}^{n-1}\left(\frac{1}{n}\sum_{l=0}^{n-1}\chi_{\omega^k}\left(\omega^l\right)a_{\omega^l}\right)\phi_{\omega^k} \\ 
 z_S^* & = & \sum_{k=0}^{n-1}b_{\omega^k}\psi_{\omega^k} \\ 
 z_T^* & = & \sum_{k=0}^{n-1}\left(\frac{1}{n}\sum_{l=0}^{n-1}\chi_{\omega^k}\left(\omega^l\right)\alpha_{\omega^l}\right)\psi_{\omega^k}. \eea
 
\section*{$\mathbb{C}(S_3)$:} 

The group $S_3=\left(1,r,r^2, \tau, \tau r, \tau r^2\right)$. It has three conjugacy classes, $[S_3]= \left([1], [r], [\tau]\right)$, where $[1]=\{1\}$, $[r] =\{r, r^2\}$ and $[\tau] = \{\tau, \tau r, \tau r^2\}$. We denote the group elements by $g$ , the elements of the group algebra and its dual by $\phi_g$ and $\psi_g$ respectively. 

The transfer matrix is given by
\bea\label{ts3} U\left(\mathbb{C}(S_3), z_S, z_T, z_S^*, z_T^*\right) & = &  \prod_p \left(\sum_{C\in [S_3]} \beta_CB_p^C\right) \nn \\ 
 & \times & \prod_j\left(\sum_{g\in S_3} b_gT_j^g\right) \prod_i \left(\sum_{R\in \textrm{IRR's of $S_3$}} a_RX_i^R\right) \prod_v A_v \eea
 with 
 \bea \label{operatorss3}A_v & = & \sum_{g\in S_3}\alpha_g~L_{i_1}^g\otimes R_{i_2}^{g^{-1}}\otimes L_{i_3}^g\otimes R_{i_4}^{g^{-1}} \\
 B_p^C & = & \sum_{\{g_i\}}\delta\left(g_1 g_2 g_3 g_4,C \right) T_{j_1}^{g_1}\otimes T_{j_2}^{g_2}\otimes T_{j_3}^{g_3^{-1}}\otimes T_{j_4}^{g_4^{-1}} \\
\label{operatorsXs3} X_i^R & = & \frac{1}{|(S_3)_R|}\sum_{g\in S_3}\chi_R(g) L_i^g.  \eea
 
The operators $T_i^g$ are projectors to the various elements $g\in S_3$. They are given by the following relations
\bea T^1 & = & \mathbbm{1} + A + Z_{11} + (Z^2)_{11} \\ 
 T^r & = & \mathbbm{1} + A + \omega^2Z_{11} + \omega (Z^2)_{11} \\
 T^{r^2} & = & \mathbbm{1} + A + \omega Z_{11} + \omega^2 (Z^2)_{11} \\
 T^\tau & = & \mathbbm{1} - A + Z_{22} + (Z^2)_{22} \\
 T^{\tau r} & = & \mathbbm{1} - A + \omega^2Z_{22} + \omega (Z^2)_{22} \\
 T^{\tau r^2} & = & \mathbbm{1} - A + \omega Z_{22} + \omega^2 (Z^2)_{22} \eea
 where 
 \bea \mathbbm{1} & = & \frac{1}{6}\left(\begin{array}{cc} 1 & 0 \\ 0 & 1\end{array}\right) ;~ A = \frac{1}{6}\left(\begin{array}{cc} 1 & 0 \\ 0 & -1\end{array}\right) \nn \\ 
   Z_{11} & = & \frac{1}{3}\left(\begin{array}{cc} Z & 0 \\ 0 & 0\end{array}\right) ;~ (Z^2)_{11} = \frac{1}{3}\left(\begin{array}{cc} Z^2 & 0 \\ 0 & 0\end{array}\right) \nn \\
   Z_{22} & = & \frac{1}{3}\left(\begin{array}{cc} 0 & 0 \\ 0 & Z\end{array}\right) ;~ (Z^2)_{22} = \frac{1}{3}\left(\begin{array}{cc} 0 & 0 \\ 0 & Z^2\end{array}\right) \eea
   where each of the matrices is divided in blocks of three by three matrices. 1 denotes the three by three identity and $Z$ is the generator of $\mathbb{Z}_3$ defined in Eq.(\ref{genzn}). 
   
   The parameters on the two sides of Eq.(\ref{ts3}) are related by
   \bea z_S & = & \beta_{[1]}\phi_1 + \beta_{[r]}\left(\phi_r + \phi_{r^2}\right) + \beta_{[\tau]}\left(\phi_\tau + \phi_{\tau r} + \phi_{\tau r^2}\right) \\
   z_T & = & \sum_{R\in S_3} a_R \sum_{g\in S_3}\chi_R(g) \phi_g \\ 
 z_S^* & = & \sum_{g\in S_3}b_g\psi_g \\
 z_T^* & = &  \sum_{g\in S_3} \alpha_g\psi_g. \eea

 The transfer matrix in Eq.(\ref{tg}) can be used to obtain the Hamiltonians through
 \beq U\left(\mathbb{C}(G), z_S, z_T, z_S^*, z_T^*\right) = e^{-H}\eeq
 where $H$ is the Hamiltonian. In the next section we will see that there are several ways of grouping terms in the transfer matrices while taking their logarithms which result in different Hamiltonians for a given set of parameters. 
 
  \section{Hamiltonians from a Deformed Transfer Matrix}
 
 The Hamiltonians can be got from the transfer matrix $U\left(\mathbb{C}(G), z_S, z_T, z_S^*, z_T^*\right)$ by taking its logarithm. There are several ways to do this and in general we can obtain many complicated models with several spins interacting. However for simplicity we will only consider four spin couplings.  Unless specified we will work only on the square lattice. However the models we define are not sensitive to the details of the lattice in the sense that the phase they describe is the same irrespective of the choice of the lattice. 

 The first two operators on the vertices and plaquettes are four spin interaction terms by construction. The operators on the links $i$, $j$ are not interaction terms between spins but magnetic field terms or perturbations~\cite{Castel1} as noted previously. These terms do not commute with the vertex and plaquette operators in general making the procedure for taking logarithms in their presence cumbersome. 
 
 We now write down exactly solvable models with vertex operators modified by the perturbations and which continue to remain in a topological phase\footnote{The transfer matrix corresponding to these modified operators is not shown explicitly here as it is a straightforward exercise with no further implications for this paper.}. This model is first written for a general group algebra $\mathbb{C}(G)$ and then we discuss the Abelian and non-Abelian cases separately. 
 
 \section*{$\mathbb{C}(G)$:}
 
The Hamiltonian is given by 
\beq\label{sidesG} H = \sum_v A_v^g + \sum_{C\in [G]}\beta_C\sum_p B_p^C \eeq
where $B_p^C$ is the projector to a conjugacy class $C$ of $G$ and is given by
\beq B_p^ C = \sum_{g\in C} B_p^g \eeq
and $\beta_C$ is a real parameter. 

The modified vertex operator $A_v^g$ is labeled by an element $g\in G$. It is given by 
\bea A_v^g & = &  L_{i_1}^g\left[\sum_{h\in G}a_h T^h_{i_1}\right]\otimes R_{i_2}^{g^{-1}}\left[\sum_{h\in G}b_h T^h_{i_2}\right]\otimes L_{i_3}^g\left[\sum_{h\in G}a_h T^h_{i_3}\right]\otimes R_{i_4}^{g^{-1}}\left[\sum_{h\in G}b_h T^h_{i_4}\right] \nn \\ & + & 
L_{i_1}^{g^{-1}}\left[\sum_{h\in G}a_h^* T^{gh}_{i_1}\right]\otimes R_{i_2}^g\left[\sum_{h\in G}b_h^* T^{hg^{-1}}_{i_2}\right]\otimes L_{i_3}^{g^{-1}}\left[\sum_{h\in G}a_h^* T^{gh}_{i_3}\right]\otimes R_{i_4}^g\left[\sum_{h\in G}b_h^* T^{hg^{-1}}_{i_4}\right] \eea
with the complex parameters $a_h$ and $b_h$ satisfying 
\bea a_{h_1}a_{h_1}^* & = & a_{h_2}a_{h_2}^*,  \forall h_1, h_2 \in G~\textrm{and}~h_1\neq h_2 \nn \\  
 b_{h_1}b_{h_1}^* & = & b_{h_2}b_{h_2}^*,  \forall h_1, h_2 \in G~\textrm{and}~h_1\neq h_2 \nn \\
 b_{gu} & = & \frac{a_{ug^{-1}}}{a_u}b_u, \forall u\in G. \eea
 
 The choice of parameters depend on the orientation of the $2D$ lattice. The oriented lattice is shown in fig(\ref{lattop}). Every time the arrow goes into the vertex we use the $b_h$ set of parameters. The $a_h$ set of parameters is used when the arrow goes away from the vertex. With this definition the model can be defined on a $2D$ lattice with an arbitrary triangulation. For these choice of parameters the vertex operators commute for adjacent vertices. 
 
 The commutation between the modified vertex operator $A_v^g$ and the plaquette operator $B_p^C$ projecting to the conjugacy class $C$ is non-trivial. For this proof we need not consider the $\sum_{h\in G}a_hT^h$ as it trivially commutes with the plaquette operator. The proof for the remaining part of the vertex operator goes as follows. 
 
\begin{figure}[h!]
  \centering
  \includegraphics[scale=1.2]{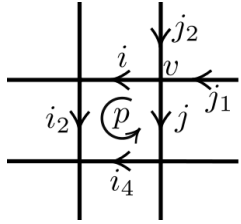}
  \caption{The oriented lattice accompanying the proof in Eq.(\ref{lhs}) and Eq.(\ref{rhs}).}\label{proof}
 \end{figure}

 Consider the plaquette $p$ and the vertex $v$ shown in oriented square lattice of fig(\ref{proof}).

 They share two links labeled $i$ and $j$ in the fig(\ref{proof}). According to the orientation of the links shown in the fig(\ref{proof}) we have
 \bea\label{lhs} A_v^gB_p^C & = & \left(L_i^g\otimes L_j^g \otimes R_{j_1}^{g^{-1}}\otimes R_{j_2}^{g^{-1}}\right)\left[\sum_{\{g_i\}}\delta\left(g_1g_2g_4g_3, C\right)T_i^{g_1}\otimes T_{i_2}^{g_2}\otimes T_j^{g_3^{-1}}\otimes T_{i_4}^{g_4^{-1}} \right] \nn \\ 
  & = & \left(\mathbbm{1}\otimes \mathbbm{1}\otimes R_{j_1}^{g^{-1}}\otimes R_{j_2}^{g^{-1}}\right)\left[\sum_{\{g_i\}}\delta\left(g_1g_2g_4g_3, C\right) L^g_iT_i^{g_1}\otimes T_{i_2}^{g_2}\otimes L^g_jT^{g_3^{-1}}_j\otimes T^{g_4^{-1}}_{i_4}\right]\eea
  and  
 \bea B_p^CA_v^g\label{rhs} & = & \left[\sum_{\{g_i\}}\delta\left(g_1g_2g_4g_3, C\right) T_i^{g_1}\otimes T_{i_2}^{g_2}\otimes T_j^{g_3^{-1}}\otimes T_{i_4}^{g_4^{-1}} \right] \left(L_i^g\otimes L_j^g \otimes R_{j_1}^{g^{-1}}\otimes R_{j_2}^{g^{-1}}\right) \nn \\ 
 & = & \left[\sum_{\{g_i\}}\delta\left(g_1g_2g_4g_3, C\right) L^g_i T_i^{g^{-1}g_1}\otimes T_{i_2}^{g_2}\otimes L_j^gT_j^{g^{-1}g_3^{-1}}\otimes T_{i_4}^{g_4^{-1}}\right]\left(\mathbbm{1}\otimes \mathbbm{1}\otimes R_{j_1}^{g^{-1}}\otimes R_{j_2}^{g^{-1}}\right) \nn \\ 
 & = & \left[\sum_{\{g_i\}}\delta\left(u_1g_2g_4u_3, C\right) L^g_i T_i^{u_1}\otimes T_{i_2}^{g_2}\otimes L_j^gT_j^{u_3}\otimes T_{i_4}^{g_4^{-1}}\right]\left(\mathbbm{1}\otimes \mathbbm{1}\otimes R_{j_1}^{g^{-1}}\otimes R_{j_2}^{g^{-1}}\right) \eea
 where we have used the relation
 \beq T^h_iL^g_i = L^g_iT^{g^{-1}h}_i. \eeq
 Thus the shifted summation in Eq.(\ref{rhs}) only shuffles the terms in the projector. This shows that the two summations in Eq.(\ref{lhs}) and Eq.(\ref{rhs}) are the same implying $A_v^g$ commutes with $B_p^C$. We can use a similar proof to show that the other three vertex operators commute with the plaquette operator. Thus the Hamiltonian in Eq.(\ref{sidesG}) is exactly solvable.
 
 Note that we have not used the full projectors for the modified vertex operators. This can be simply understood by looking at the identity term in the expression for the projectors. The identity term coupled with the perturbations is like adding magnetic fields to the system. This will take us out of the exactly solvable regime we are interested in and also we will no longer remain in a topological phase. Thus we avoid full projectors for the modified vertex operators. 
 
 We can write down a similar model where we modify the plaquette operator with the $X_i^R$ operators defined in Eq.(\ref{operatorsXG}). In the examples to follow we will write down these models for the $\mathbb{C}(\mathbb{Z}_n)$ case alone. The corresponding models for the non-Abelian cases are more cumbersome and so we do not show them here. Before we write down the specific examples let us look at the deformed quantum double algebra.
 
 \subsection*{Deformed quantum double algebra}
 
 Consider the product of two deformed vertex operators $A_v^{g_1}$ and $A_v^{g_2}$. It is given by
 \beq A_v^{g_1}A_v^{g_2} = A_v^{g_1g_2} M(g_1, g_2), \eeq
 with 
 \beq M(g_1, g_2)  = \left[\sum_h a_{g_2h}T^h_{i_1}\right]\otimes\left[\sum_h b_{hg_2^{-1}}T^h_{i_2}\right]\left[\sum_h a_{g_2h}T^h_{i_3}\right]\otimes\left[\sum_h b_{hg_2^{-1}}T^h_{i_4}\right].\eeq
 We can now check associativity of the product by comparing $(A_v^{g_1}A_v^{g_2})A_v^{g_3}$ and $A_v^{g_1}(A_v^{g_2}A_v^{g_3})$ which requires
 \beq \label{mc}M(g_1g_2, g_3)M^{g_3}(g_1, g_2) = M(g_1, g_2g_3)M(g_2, g_3),\eeq
 which is seen to be true by using the expression for $M(g_1, g_2)$ and from
 \beq M^g_3(g_1, g_2) = \left[\sum_h a_{g_2g_3h}T^h_{i_1}\right]\otimes\left[\sum_h b_{hg_3^{-1}g_2^{-1}}T^h_{i_2}\right]\left[\sum_h a_{g_2g_3h}T^h_{i_3}\right]\otimes\left[\sum_h b_{hg_3^{-1}g_2^{-1}}T^h_{i_4}\right].\eeq
 The quantum double algebra given by 
 \beq A^g_vB_p^h = B^{ghg^{-1}}_pA_v^g,\eeq
 is unchanged but the product in the algebra is changed to
 \beq \left(A_v^{g_1}B_p^{h_1}\right)\left(A_v^{g_2}B_p^{h_2}\right) = \delta_{h_2, g_2^{-1}h_1g_2}A_v^{g_1g_2}M(g_1, g_2) B_p^{h_2}.\eeq
 
 This algebra is similar to quasi quantum doubles as seen in~\cite{Bais1} where the deformation occurs through a 2-cocycle. Such algebras also have a non-coassociative coproduct with an associator given by a 3-cocycle related to the 2-cocycle by a slant product~\cite{Bais1}. The condition in Eq.(\ref{mc}) is similar to the one obtained in studying obstructions to implementations of global symmetries on physical systems as used in analysis of symmetry protected topological phases~\cite{nayakelse}. Note that we have not studied the co-algebra structure of these operators here. It is not necessary for what is to follow and we stop with the above crucial remarks.
  
 \section*{$\mathbb{C}(\mathbb{Z}_2)$:} 
 
 The Hamiltonian in this case is given by
 \beq\label{sides2v} H = \sum_v A_v' + \sum_p\left(\beta_1 B_p^1 + \beta_{-1}B_p^{-1}\right) \eeq
 with $A_v'$ given by 
 \bea A_v'& = & \frac{1}{(b_1b_{-1})^2}\left[\sigma_{i_1}^x\left(b_1T^1_{i_1} + b_{-1}T^{-1}_{i_1}\right)\otimes\sigma_{i_2}^x\left(b_1T^1_{i_2} + b_{-1}T^{-1}_{i_2}\right)\otimes\sigma_{i_3}^x\left(b_1T^1_{i_3} + b_{-1}T^{-1}_{i_3}\right)\otimes\sigma_{i_4}^x\left(b_1T^1_{i_4} + b_{-1}T^{-1}_{i_4}\right) \right. \nn \\ & + & \left. \sigma_{i_1}^x\left(b_1^*T^{-1}_{i_1} + b_{-1}^*T^1_{i_1}\right)\otimes\sigma_{i_2}^x\left(b_1^*T^{-1}_{i_2} + b_{-1}^*T^1_{i_2}\right)\otimes\sigma_{i_3}^x\left(b_1^*T^{-1}_{i_3} + b_{-1}^*T^1_{i_3}\right)\otimes\sigma_{i_4}^x\left(b_1^*T^{-1}_{i_4} + b_{-1}^*T^1_{i_4}\right)\right] \eea
 where the complex parameters $b_{\pm 1}$ have the same modulus. 
 
 It is easy to check that the model is exactly solvable as the terms commute with each other. To compute the spectrum of the Hamiltonian we need to find the eigenvalues of the modified vertex operator $A_v'$ and the plaquette operators. Each plaquette operator is a sum of two projectors and so the eigenvalues of that operator depends on the magnitude of $\beta_{\pm 1}$. The eigenvlaues of $A_v'$ are given by $\pm 1$.

 The dual of the above model in Eq.(\ref{sides2v}) is given by the following Hamiltonian
 \beq\label{sides2} H = \sum_v\left(\alpha_1 A_v^1 + \alpha_{-1}A_v^{-1}\right) + \sum_p B_p' \eeq
 where
 \beq B_p' = \frac{1}{(a_1a_{-1})^2}\left(B_p^1 - B_p^{-1}\right)\prod_{i_k\in\partial p, k=1}^4\left(a_1 X_{i_k}^1 + a_{-1}X_{i_k}^{-1}\right) + h.c. \eeq
 with $\partial p$ being the perimeter of the plaquette $p$. This model can be defined on any two dimensional lattice for an arbitrary triangulation. On the square lattice this plaquette operator continues to be a four spin interaction term as in the usual quantum double model. It is easy to check that for two adjacent plaquettes $p_1$ and $p_2$, these operators commute and that $(B_p')^2 = 1$. This operator clearly commutes with the vertex operators as can be easily seen from the expression for $B_p'$. Thus the Hamiltonian in Eq.(\ref{sides2}) continues to be an exactly solvable model.    

The Hamiltonian in Eq.(\ref{sides2}) is got by the following choice of the parameters in the transfer matrix of Eq.(\ref{tz2}):
 \bea 
 z_S^* & = & \psi_1 + \psi_{-1} \\ 
z_T^* & = & \left(\frac{e^{\alpha_1}+e^{\alpha_{-1}}}{2}\right)\psi_1 + \left(\frac{e^{\alpha_1} - e^{\alpha_{-1}}}{2}\right)\psi_{-1}.\eea
 The values of the other two parameters $z_S$ and $z_T$ can be found but after a cumbersome calculation involving ten variables. We do not show this computation here but just remark that it is possible to find these parameters as well. Though this computation may seem irrelevant it is an important calculation to show that the semion model cannot be obtained from the transfer matrix in Eq.(\ref{tz2}). We will make more comments about this in the remarks section. 
 
In general it is important to consider the orientation of the two dimensional triangulated lattice. There are orientations for the links and the plaquettes of this lattice. These are shown in fig(\ref{lattop}). For the $\mathbb{C}(\mathbb{Z}_2)$ case the orientations are not important as the inverse of the group elements in this case is the same as the group element itself. 

%We now consider the $\mathbb{C}(\mathbb{Z}_3)$ model.

%The Hamiltonian in this case is given by 
%\beq\label{sides3} H = \sum_v\left(\alpha_1 A_v^1 + \alpha_\omega A_v^\omega + \alpha_{\omega^2}A_v^{\omega^2}\right) + \sum_p B_p' \eeq
%where $B_p'$ is defined by
%\bea B_p' & = & \frac{1}{(a_1a_\omega a_{\omega^2})^{\frac{2}{3}}(b_1b_\omega b_{\omega^2})^{\frac{2}{3}}}\left(B_p^1 + \omega^2B_p^\omega + \omega B_p^{\omega^2}\right)\nn \\ 
%& \times & \prod_{i_{k,k'}\in\partial p, k=(1,2), k'=(3,4)}\left(a_1 X_{i_k}^1 + a_\omega X_{i_k}^\omega + a_{\omega^2} X_{i_k}^{\omega^2}\right)\left(b_1 X_{i_k'}^1 + b_\omega X_{i_k'}^\omega + b_{\omega^2} X_{i_k'}^{\omega^2}\right) \eea
%with $b_\omega = \frac{a_{\omega^2}}{a_1}b_1, b_{\omega^2} = \frac{a_{\omega^2}}{a_\omega}b_1$. 

%It is easy to check that $B_{p_1}'$ and $B_{p_2}'$ commutes for adjacent plaquettes $p_1$ and $p_2$ and that $(B_p')^3 = 1$. It also commutes with the vertex operator in a trivial way thus keeping the Hamiltonian in Eq.(\ref{sides3}) an exactly solvable one. 
\section*{$\mathbb{C}(\mathbb{Z}_n)$:}

The Hamiltonian in this case is given by 
\beq\label{sidesnv} H = \sum_v \left(A_v' + h.c.\right) + \sum_p \left(\sum_{k=0}^{n-1}\beta_{\omega^k}B_p^{\omega^k}\right) \eeq
with $A_v'$ given by 
\bea A_v' & = & \frac{1}{\left(\prod_{k=0}^{n-1}a_{\omega^k}\right)^{\frac{2}{n}}\left(\prod_{k=0}^{n-1}b_{\omega^k}\right)^{\frac{2}{n}}}|\left(\sum_{k=0}^{n-1}\omega^{n-k}A_v^{\omega^k}\right) \nn \\ & \times & \prod_{i_{k,k'}\partial p, k=(1,2), k'=(3,4)}\left(\sum_{j=0}^{n-1}a_{\omega^j}T_{i_k}^{\omega^j}\right)\left(\sum_{j=0}^{n-1}b_{\omega^j}T_{i_k'}^{\omega^j}\right) \eea
where the parameters $a_{\omega^k}$ and $b_{\omega^k}$ satisfy $a_{\omega^{k-1}}b_{\omega^k} = a_{\omega^k}b_{\omega^{k+1}}$ for $k\in \left(0,\cdots, n-1\right)$ and $a_{\omega^{k_1}}a_{\omega^{k_1}}^* = a_{\omega^{k_2}}a_{\omega^{k_2}}^*$, $b_{\omega^{k_1}}b_{\omega^{k_1}}^* = b_{\omega^{k_2}}b_{\omega^{k_2}}^*$ for $k_1\neq k_2$. These operators commute for adjacent vertices when the parameters satisfy these conditions. It is also easy to check that $(A_v')^n=1$ and hence it has the nth roots of unity as its eigenvalues. As the hermitian conjugate is added to $A_v'$ the eigenvalues are real. 

The dual version of the model in Eq.(\ref{sidesnv}) is given by the Hamiltonian
\beq\label{sidesn} H = \sum_v\left(\sum_{k=0}^{n-1} \alpha_{\omega^k}A_v^{\omega^k}\right) 
+\sum_p \left(B_p'+h.c.\right) \eeq
where $B_p'$ is given by
\bea B_p' & = & \frac{1}{\left(\prod_{k=0}^{n-1}a_{\omega^k}\right)^{\frac{2}{n}}\left(\prod_{k=0}^{n-1}b_{\omega^k}\right)^{\frac{2}{n}}}\left(\sum_{k=0}^{n-1} \omega^{n-k} B_p^{\omega^k}\right) \nn \\ 
& \times & \prod_{i_{k,k'}\in\partial p, k=(1,2), k'=(3,4)} \left(\sum_{j=0}^{n-1} a_{\omega^j}X_{i_k}^{\omega^j}\right)\left(\sum_{j=0}^{n-1} b_{\omega^j}X_{i_k'}^{\omega^j}\right) \eea
with the solutions for the $b$'s given  by the relations $ a_{\omega^{k-1}}b_{\omega^k} = a_{\omega^k}b_{\omega^{k+1}}, k\in (0\cdots n-1)$. The modulus of these parameters are the same as in their dual version. These operators commute for adjacent plaquettes and $(B_p')^n = 1$. 

\section*{$\mathbb{C}(S_3)$:}

The group $S_3$ has six elements given by $S_3 = \left(1, r, r^2, \tau, \tau r, \tau r^2\right)$ with $r^3=\tau^2=1$ and $r\tau =\tau r^2$. We choose $g$ to belong to a normal subgroup $N$ of $S_3$. The only normal subgroup of $S_3$ is given by $N= \left(1, r, r^2\right)$. Thus the modified vertex operator has either $g=r$ or $g=r^2$. There are three conjugacy classes for $S_3$ given by $[S_3] = \left(\{1\}, \{r, r^2\}, \{\tau, \tau r, \tau r^2\}\right)$. 

The Hamiltonian is given by 
\beq\label{sides3v} H = \sum_v \left(A_v^r + h.c.\right) + \sum_p\sum_{C\in [S_3]}\beta_C\left(\sum_{k\in C}B_p^k \right) \eeq
with $A_v^r$ given by 
\beq A_v^r  =  L^r_{i_1}\left(\sum_{h\in S_3} a_hT^h\right)\otimes R^{r^2}_{i_2}\left(\sum_{h\in S_3} b_hT_{i_2}^h\right)\otimes L^r_{i_3}\left(\sum_{h\in S_3} a_hT_{i_3}^h\right)\otimes R^{r^2}_{i_4}\left(\sum_{h\in S_3} b_hT_{i_4}^h\right) \eeq 
with the parameters satisfying the following 
\bea b_r & = & \frac{a_{r^2}}{a_1}b_1;~ b_{r^2} = \frac{a_{r^2}}{a_r}b_1 \nn \\ 
 b_{\tau r^2} & = & \frac{a_{\tau r^2}}{a_\tau}b_\tau;~ b_{\tau r} = \frac{a_{\tau r}}{a_\tau}b_\tau \eea
 and they have the same modulus. 
 
 The eigenvalues of $A_v^r$ are found by computing its cube as $r^3=1$. This gives the eigenvalues as $\left(a_1a_ra_{r^2}\right)^2\left(b_1b_rb_{r^2}\right)^2$ and $\left(a_\tau a_{\tau r}a_{\tau r^2}\right)^2\left(b_\tau b_{\tau r}b_{\tau r^2}\right)^2$. 
 
 For the chosen parameters the model is solvable as all the terms commute with each other. 
 
 We can define exactly solvable models by choosing $g$ to be either $\tau$, $\tau r$, or $\tau r^2$. These models are in another phase which we do not discuss here. 

 In the next section we find out the quantum phases in which these models live in by studying their excitations.

\section{Excitations, Statistics, Fusion Rules}

We find out the quantum phase for the three examples considered.

\section*{$\mathbb{C}(\mathbb{Z}_2)$:}

The ground states $|G\rangle$ for the Hamiltonian in Eq.(\ref{sides2}) is given by the following conditions
\beq A_v^1|G\rangle = |G\rangle, A_v^{-1}|G\rangle = 0, B_p'|G\rangle = -|G\rangle;~ \forall ~v, p.\eeq To validate these conditions we assume the relative magnitudes of the parameters in the Hamiltonian in Eq.(\ref{sides2}) according to the conditions stated. 

For the $\mathbb{C}(\mathbb{Z}_2)$ case we have the following operators create the flux (violations of the plaquette operator conditions) and charge (violation of the vertex operator conditions).
\bea\label{excite2} V_{\gamma^*}^f & = & \prod_{i\in \gamma^*} \sigma_i^x \\
V_\gamma^c & = & \frac{1}{(n_1n_{-1})^{\frac{1}{2}}}\prod_{i\in\gamma} \sigma_i^z\prod_{j\in\gamma}\left(n_1X_j^1 + n_{-1}X_j^{-1}\right) \eea
where $\gamma$ and $\gamma^*$ are strings in the direct and dual lattice respectively. The parameters $n_1$ and $n_{-1}$ satisfy $n_1 = \frac{a_1}{a_{-1}}n_{-1}$. This condition makes the operator creating the flux independent of $n_{\pm 1}$. 

The dyonic excitations is created by applying both these strings on the dual and direct lattice respectively.  

The statistics of these excitations of these particles can easily be found out using the operators in Eq.(\ref{excite2}). The self statistics of fluxes are trivial as the operator creating the flux is the same as in the toric code case. Hence they remain bosonic. The self statistics of charges is also seen to be bosonic as
\beq \frac{1}{n_1n_{-1}}\sigma^z_i\left(n_1 X_i^1 + n_{-1}X_i^{-1}\right)\sigma^z_i\left(n_1 X_i^1 + n_{-1}X_i^{-1}\right) = 1.\eeq

The mutual statistics between the charges and the fluxes is clearly fermionic as can be trivially seen from the expressions of the operators creating the charges in Eq.(\ref{excite2}). 
 
 For convenience we now label the particles in this model as 1, $\tilde{e}$, $m$ and $\epsilon$. 1 is the vacuum, $\tilde{e}$ is the charge, $m$ is the flux and $\epsilon$ is the dyonic excitation. The fusion rules can be obtained from the algebra of the orators creating them in Eq.(\ref{excite2}), and they are found to be
 \bea\label{fusion2} \tilde{e}\times\tilde{e} & = & m\times m= \epsilon\times \epsilon = 1 \nn \\ 
 \tilde{e}\times m & = & \epsilon \eea
 and the remaining fusion rules involving the fusion of the vacuum are all trivial. Thus we find the fusion rules to be the same as the ones for the $\mathbb{C}(\mathbb{Z}_2)$ anyon model. 
 
 \section*{$\mathbb{C}(\mathbb{Z}_n)$:}
 
 For the $\mathbb{C}(\mathbb{Z}_n)$ case, the Hamiltonian is given by Eq.(\ref{sidesn}). The conditions for the ground states are similar to the $n=2$ case. Again we assume the parameters to be such that these conditions hold. As in the $n=2$ case the fluxes do not change but the operators creating the charges have to be modified. There are $n-1$ charges and we denote them by $e^{n-k}$ with $k\in \left(1,\cdots, n-1\right)$. The operators creating them are given by 
 \beq \label{chargeN}V_\gamma^{e^{n-k}} = Z^{n-k}\left(\sum_{j=0}^{n-1}m_{(n-k)\omega^j}X^{\omega^j}\right) \eeq
 where $\gamma$ is a direct triangle~\cite{Del1} and the parameters $m_{(n-k)\omega^j}$ satisfy 
 \beq \label{recurN}a_{\omega^{j-k}}m_{(n-k)\omega^j} = m_{(n-k)\omega^{j-n+1}}a_{\omega^j}\eeq 
 for $j\in \left(0,\cdots, n-1\right)$. 
 At this point there seems to be one free parameter among the $m_{(n-k)\omega^j}$'s for a given charge $e^{n-k}$. The remaining are found in terms of this single ``free'' parameter using the recursion relations in Eq.(\ref{recurN}). However this is not there once we consider the fusion rules. They consistently fix all the parameters $m_{(n-k)\omega^j}$ in the string.
 
 The string operators for the fluxes are the same as in the $\mathbb{C}(\mathbb{Z}_n)$ quantum double model and are given by
 \beq\label{fluxN} V_{\gamma^*}^{f^k} =  X_j^k \eeq
 where $\gamma^*$ is a dual triangle and $j$ is the label for this edge. 
 
 It is easy to check using the string operators in Eq.(\ref{chargeN}) and Eq.(\ref{fluxN}) that the statistics of these particles are the same as in a $\mathbb{Z}_n$ anyon model. The charges and fluxes have bosonic self-statistics. The mutual statistics are not trivial and they are given by
 \beq e^{n-k_1}\times f^{k_2} = \omega^{(n-k_1)k_2}f^{k_2}\times e^{n-k_1}. \eeq
 
 The fusion rules of the fluxes do not change. The fusion rules of the charges are the same when we constrain the free parameters of the parameters $m_{(n-k)\omega^j}$. This constrain is given by
 \beq m_{(n-k_1)\omega^{j-k_2}}m_{(n-k_2)\omega^j} = m_{(n-(k_1+k_2))\omega^j} \eeq
 for $k_1$, $k_2$ belong to $(1,\cdots, n-1)$ and $j\in \left(0,\cdots, n-1\right)$. 
 This relation comes when we impose the $\mathbb{Z}_n$ fusion rules for the charges for the operators creating the charges in Eq.(\ref{chargeN}) which is 
 \beq e^{n-k_1}\times e^{n-k_2} = e^{n-(k_1+k_2)}.\eeq
 These fix all the $m_{(n-k)\omega^j}$ parameters unambiguously. 
 
 As we have shown the existence of all the charges and fluxes of the $\mathbb{C}(\mathbb{Z}_n)$ quantum double model exist in our model they are in the same phase. 
 
 A similar construction can be carried out when the vertex operator is modified and the flux operator is left unchanged as given in Eq.(\ref{sidesnv}). This model is also in the same phase as the $\mathbb{C}(\mathbb{Z}_n)$ quantum double phase. 
 
 \section*{$\mathbb{C}(S_3)$:}
 
 The quantum phase of this case is not in the phase of the corresponding quantum double model. This is because some of the particles in the quantum double phase condense and some others do not exist. We will write down which ones condense and which ones do not exist for the $S_3$ example in what follows. We will work with the vertex version of the model given by Eq.(\ref{sides3v}). 
 
 The $S_3$ anyon model consists of eight excitations including the vacuum~\cite{Overbosch}. They are given by $$1, A(e), B(e),  Id(r), Id(\tau), r1(r), r2(r), A(\tau).$$ 1 is the vacuum, $A(e)$ and $B(e)$ are the Abelian and non-Abelian charges respectively. $Id(r)$ and $Id(\tau)$ are the non-Abelian fluxes in the model. $r1(r)$, $r2(r)$ and $A(\tau)$ are the dyonic excitations.
 
 In the notation of~\cite{Overbosch} these excitations are given by $$1, A, J^w, J^x, K^a, J^y, J^z, K^b.$$ We will switch to this notation when we discuss the fusion rules in this model. 
 
 The operators creating these excitations, called ribbon operators, can be found in~\cite{Del1, LevinNA}. We do not write all these operators here. We will only write down operators which will be modified in our model. 
 
 The Abelian charge $A(e)$ condenses as the ribbon operator creating this excitation commutes with both the plaquette and the modified vertex operators. The ribbon operator creating the non-Abelian charge $B(e)$ is not modified. 
 
 The non-Abelian flux $Id(r)$ is created by the following modified operators.
 \bea \label{modfl3}\tilde{F}_{\gamma;~Id(r)}^{(1,1)(1,1)} & = & F_{\gamma;~Id(r)}^{(1,1)(1,1)}\left(\sum_{h\in S_3} n_hT^h\right) \nn \\
 \tilde{F}_{\gamma;~Id(r)}^{(2,1)(2,1)} & = & F_{\gamma;~Id(r)}^{(2,1)(2,1)}\left(\sum_{h\in S_3} m_hT^h\right) \eea
where the parameters are given by 
\bea n_r & = & \frac{a_r}{a_1}n_1;~ n_{r^2} = \frac{a_{r^2}}{a_1}n_1 \\
n_{\tau r} & = & \frac{a_{\tau r}}{a_\tau}n_\tau;~ n_{\tau r^2} = \frac{a_{\tau r^2}}{a_\tau}n_\tau \eea
and
\bea m_{r^2} & = & \frac{a_r}{a_1}m_1;~ m_r = \frac{a_r}{a_{r^2}}m_1 \\
m_{\tau r^2} & = & \frac{a_{\tau r}}{a_\tau}m_\tau;~ m_{\tau r} = \frac{a_{\tau r}}{a_{\tau r^2}}m_\tau. \eea

At this point $n_1$, $n_\tau$, $m_1$ and $m_\tau$ seem to be free parameters. However the ribbon operators should not depend on these free particles as this will indefinitely increase the number of particles. We will see that the fusion rules fix these parameters unambiguously.  

The non-Abelian flux $Id(\tau)$ does not exist in this case as if we attempt to modify the corresponding ribbon operator to commute with the modified vertex operator we will be forced to constrain the parameters $a_h$ which eventually makes the model trivial in the sense they remove the additional qudit terms on the sides. Hence this changes the model and the phase. So we conclude that in this phase this flux excitation does not exist or they create states which cannot be expanded in the basis of the excitations of the condensed model. This is not surprising as even in the $\mathbb{C}(S_3)$ quantum double model we can write operators which after acting on the ground states create states that cannot be expanded in the basis of the eight excitations of the $S_3$ anyon model. One such operator is given by the following matrix
\beq O = \left(\begin{array}{cc} 0 & Z \\ 0 & 0\end{array}\right).\eeq 
It can be easily checked that this matrix does not commute with both the plaquette and vertex operators of the quantum double model of $\mathbb{C}(S_3)$. It can also be checked that this matrix cannot be expanded in terms of the matrices creating the eight excitations of the $\mathbb{C}(S_3)$ quantum double model. The operator creating the flux excitation $Id(\tau)$ faces a similar situation in our model. 

The non-existence of this flux also means that the dyonic excitation $A(\tau)$ does not exist in this model. 

On the other hand the dyonic excitations given by $r1(r)$ and $r2(r)$ do exist in this model and they are created by the modified operators creating flux excitations and the unchanged charge operators.

We now fix the remaining ``free'' parameters $n_1$, $n_\tau$, $m_1$, $m_\tau$ using appropriate fusion rules. The fusion rules of the $S_3$ anyon model can be found in~\cite{Overbosch}. As we do not have the excitations denoted by $K^a$ and $K^b$ we can ignore their fusion rules. They are decoupled from the fusion rules of the other particles. The fusion rule which will be useful to fix the remaining parameters is given by
\beq J^x\times J^x = 1 + A + J^x.\eeq
The operators creating the excitation $A$ are not modified in our model and hence we cannot use this channel to fix the remaining parameters. We use the other two channels instead. So we have
\beq \tilde{F}_{\gamma;~Id(r)}^{(1,1)(1,1)}\times \tilde{F}_{\gamma;~Id(r)}^{(2,1)(2,1)}  = 1.\eeq
This gives us the conditions 
\beq m_1n_1 = \frac{a_1}{a_r};~ m_\tau n_\tau = \frac{a_\tau}{a_{\tau r}}. \eeq

The second channel is obtained by
\beq  \tilde{F}_{\gamma;~Id(r)}^{(1,1)(1,1)}\times \tilde{F}_{\gamma;~Id(r)}^{(1,1)(1,1)} = \tilde{F}_{\gamma;~Id(r)}^{(2,1)(2,1)} \eeq
which gives us the conditions
\beq m_1 = \frac{a_{r^2}}{a_1}n_1^2;~ m_\tau = \frac{a_{\tau r^2}}{a_\tau}n_\tau^2.\eeq
These conditions unambiguously fix all the parameters in the ribbon operators of Eq.(\ref{modfl3}). 

Thus we have seen that this model is in a quantum phase different from the $\mathbb{C}(S_3)$ quantum double model. 

\section{The $\mathbb{Z}_2$ - Double Semionic Phase from the Transfer Matrix Picture}

The transfer matrices considered so far were made up of two kinds of operators, the plaquette and vertex operators, given by $B_p^C$ and $A_v$, which coupled four qudits and the single qudit operators, given by $X_j^R$ and $T^g_i$. Such operators arise in a transfer matrix parametrized by the centers of the algebra $\mathcal{A}$ and its dual $\mathcal{A}^*$. We denoted these transfer matrices as $$U\left(\mathcal{A}, z_S, z_T, z_S^*, z_T^*\right)$$
where $\mathcal{A}$ was taken to be the  group algebra $\mathbb{C}(G)$. The parameters $z_S$, $z_T$, $z_S^*$ and $z_T^*$ are the centers of the algebra and its dual respectively. In the pictorial language of~\cite{QDMST} these parameters were associated to the spacelike plaquettes, timelike plaquettes, spacelike links and timelike links respectively. 

 These are the operators which result in the transfer matrix of a lattice gauge theory and we studied the quantum phases we can produce using the Hamiltonians from these transfer matrices. In particular we obtained the quantum double phase of Abelian gauge groups from non-trivial Hamiltonians constructed by modifying these transfer matrices. For example in the $\mathbb{Z}_2$ case we obtained the toric code phase. However it is not possible to obtain the double semion phase from this modified transfer matrix. In order to obtain this we need to consider more general transfer matrices made up of other kinds of operators. Such operators can be obtained by considering more general transfer matrices parametrized by the centers of not just the algebra $\mathcal{A}$ but by two copies of the algebra $\mathcal{A}\otimes\mathcal{A}$ and its dual which result in operators coupling two qudits. A transfer matrix of this sort can be formally written as $$U\left(\mathcal{A}, z_S, z_T, z_S^*, z_T^*, z_{\textrm{vol}}, z_{\textrm{ver}}^*\right)$$ where $z_{\textrm{vol}}$ and $z_{\textrm{ver}}^*$ are the centers of $\mathcal{A}\otimes\mathcal{A}$ and its dual respectively. In the language of the pictorial formalism introduced in~\cite{QDMST}, these parameters are associated to the volumes and the vertices of the triangulated three dimensional lattice respectively. We will not go into the details of this pictorial formalism in this paper and only show the Hamiltonians that we can obtain from such a transfer matrix. 
 
It is possible to build such transfer matrices made up of the usual four qudit plaquette and vertex operators as well as two and three qudit operators which are combinations of $X_j^R$ and $T^g_i$, acting on different links, coupled together. Such considerations will lead to both the toric code phase and the double semion phase as we shall see now.  We note that such transfer matrices are not the transfer matrices of the usual lattice gauge theory based on discrete groups. We only work with the case of $G=\mathbb{Z}_2$ in what follows.  

\begin{figure}[h!]
  \centering
  \includegraphics[scale=1.2]{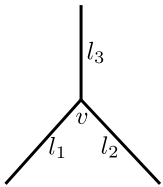}
  \caption{The vertex on which the modified vertex operator acts.}\label{vertex}
 \end{figure}

We will work on a lattice where there are vertices that have valency three. The modified vertex operator acting on this vertex is shown in fig(\ref{vertex}). 

In general the lattice is oriented but as we are working with the $\mathbb{Z}_2$ case in this section we ignore the orientation. The operators coupling two links together are given by $$ b_{11} T_{l_1}^1T_{l_2}^1 + b_{12} T_{l_1}^1T_{l_2}^{-1} + b_{13} T_{l_1}^{-1}T_{l_2}^1 + b_{14} T_{l_1}^{-1}T_{l_2}^{-1}$$, $$ b_{21} T_{l_2}^1T_{l_3}^1 + b_{22} T_{l_2}^1T_{l_3}^{-1} + b_{23} T_{l_2}^{-1}T_{l_3}^1 + b_{24} T_{l_2}^{-1}T_{l_3}^{-1}$$ and $$ b_{31} T_{l_3}^1T_{l_1}^1 + b_{32} T_{l_3}^1T_{l_1}^{-1} + b_{33} T_{l_3}^{-1}T_{l_1}^1 + b_{34} T_{l_3}^{-1}T_{l_1}^{-1}.$$ Using these three two link operators we can write down the modified vertex operator as 
\bea\label{z2d} A_v^{-1} & = &  X_{l_1}X_{l_2}X_{l_3} \nn \\
& \times & \left(a_1 T_{l_1}^1T_{l_2}^1T_{l_3}^1 + a_2 T_{l_1}^1T_{l_2}^1T_{l_3}^{-1} + a_3 T_{l_1}^1T_{l_2}^{-1}T_{l_3}^1 + a_4 T_{l_1}^1T_{l_2}^{-1}T_{l_3}^{-1} \right. \nn\\ &+& \left. a_5 T_{l_1}^{-1}T_{l_2}^1T_{l_3}^1 + a_6 T_{l_1}^{-1}T_{l_2}^1T_{l_3}^{-1} + a_7 T_{l_1}^{-1}T_{l_2}^{-1}T_{l_3}^1 + a_8 T_{l_1}^{-1}T_{l_2}^{-1}T_{l_3}^{-1}\right). \eea

The relation between $a$'s and $b$'s are given by 
\bea a_1 & = & b_{11}b_{21}b_{31},~ a_2  = b_{11}b_{22}b_{33},~ a_3  = b_{12}b_{23}b_{31},~ a_4  = b_{12}b_{24}b_{33} \nn \\ 
a_5  & = & b_{13}b_{21}b_{32},~ a_6  = b_{13}b_{22}b_{34},~ a_7  = b_{14}b_{23}b_{32},~ a_8  = b_{14}b_{24}b_{34}.\eea

For adjacent vertex operators to commute the parameters have to satisfy the following relation
\beq \frac{a_5}{a_1} = \frac{a_6}{a_2}=\frac{a_7}{a_3}=\frac{a_8}{a_4}.\eeq
Thus there are four free parameters given by $a_1$, $a_2$, $a_3$ and $a_4$. In order to have $(A_v^{-1})^2=1$ we further restrict them with the relation $a_1a_4 = a_2a_3$. Thus there are just three free parameters now. The plaquette operator is unchanged and is that of the toric code. 

\begin{figure}[h!]
  \centering
  \includegraphics[scale=1.2]{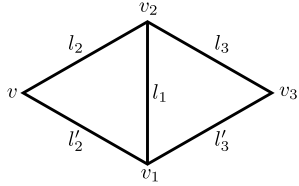}
  \caption{The link $l_1$ and the two plaquettes adjacent to it.}\label{triangle}
 \end{figure}

We now obtain the excitations in this model. The operator creating the vertex excitations that is the charges is unchanged with respect to the usual toric code. The change is seen in the operators creating the fluxes. They have to be modified in order for them to commute with the modified vertex operator given in Eq.(\ref{z2d}). Consider the following operator creating the flux on the two plaquettes adjacent to link $l_1$ as shown in the fig(\ref{triangle}). 

\bea\label{f2d} F_{l_1} & = & X_{l_1} \nn \\
& \times & \left(n_1 T_{l_1}^1T_{l_2}^1T_{l_3}^1 + n_2 T_{l_1}^1T_{l_2}^1T_{l_3}^{-1} + n_3 T_{l_1}^1T_{l_2}^{-1}T_{l_3}^1 + n_4 T_{l_1}^1T_{l_2}^{-1}T_{l_3}^{-1} \right. \nn\\ &+& \left. n_5 T_{l_1}^{-1}T_{l_2}^1T_{l_3}^1 + n_6 T_{l_1}^{-1}T_{l_2}^1T_{l_3}^{-1} + n_7 T_{l_1}^{-1}T_{l_2}^{-1}T_{l_3}^1 + n_8 T_{l_1}^{-1}T_{l_2}^{-1}T_{l_3}^{-1}\right) \nn \\ 
&\times & \left(n_1 T_{l_1}^1T_{l_2'}^1T_{l_3'}^1 + n_2 T_{l_1}^1T_{l_2'}^1T_{l_3'}^{-1} + n_3 T_{l_1}^1T_{l_2'}^{-1}T_{l_3'}^1 + n_4 T_{l_1}^1T_{l_2'}^{-1}T_{l_3'}^{-1} \right. \nn\\ &+& \left. n_5 T_{l_1}^{-1}T_{l_2'}^1T_{l_3'}^1 + n_6 T_{l_1}^{-1}T_{l_2'}^1T_{l_3'}^{-1} + n_7 T_{l_1}^{-1}T_{l_2'}^{-1}T_{l_3'}^1 + n_8 T_{l_1}^{-1}T_{l_2'}^{-1}T_{l_3'}^{-1}\right).\eea

For this operator to commute with the vertices $v_1$ and $v_2$ we have the following relations
\beq n_5  =  n_1 k,~ n_6 = n_2 k,~ n_7 = n_2 k,~ n_8 = n_1 k, \eeq
with $k = \left(\frac{a_5}{a_1}\right)^{\frac{1}{2}}$. The relation between $n_1$ and $n_2$ is fixed by the condition that $F_{l_1}$ commutes with the vertex operators associated to $v$ and $v_3$. This is given by 
\beq n_1^2 = n_2^2.\eeq
We now study the two cases separately. When $n_1=n_2$ we have the following form for the operator $F_{l_1}$ given by
\beq\label{f1} F_{l_1} = n_1^2 X_{l_1}\left[\left(\frac{1+k^2}{2}\right) + \left(\frac{1-k^2}{2}\right) Z_{l_1}\right] \eeq 
where $X$ and $Z$ are Pauli matrices acting on the link $l_1$. This string operator cannot create the semion phase and hence remains in the toric code phase. This can be seen by computing the fusion rules and the braiding statistics. The fusion rule for this operator is given by 
\beq F_{l_1}\times F_{l_1} = n_1^4k^2.\eeq
By choosing $n_1 = \frac{1}{k^{\frac{1}{2}}}$ we obtain the fusion rules for the toric code phase.  

\begin{figure}[h!]
  \centering
  \includegraphics[scale=1.2]{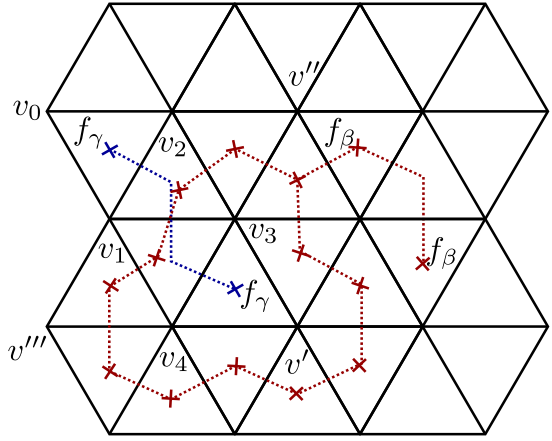}
  \caption{The braiding of the strings creating the modified fluxes.}\label{braiding}
 \end{figure}

Let us work on the hexagonal lattice as shown in fig(\ref{braiding}) for computing the braiding rules. We consider two sets of fluxes created by the operators $F_\gamma$ and $F_\beta$ respectively. The braiding is carried out as shown in fig(\ref{braiding})

and is given by 
\bea\label{gb} F_\gamma F_\beta' & = & X_{12}X_{13}X_{43}X_{23}X_{13}X_{14} \nn \\
& \times & nT^{23}_{0,2,3} nT^{13}_{0,1,3} nT^{23, 13}_{2, 3, 4} nT^{13, 14}_{2, 1, 4} nT^{13, 14}_{2, 1, 4} nT^{13}_{1, 3, v'} nT^{14}_{1, 4, v'} \nn \\ 
& \times & nT_{1, 3, v^{''}} nT_{1, 2, v^{''}} nT_{2, 1, 4} nT_{2, 3, 4} nT_{v^{'''}, 1, 3} nT_{v^{'''}, 4, 3} \eea
where $F_\beta'$ is the part of $F_\beta$ that intersects $F_\gamma$. 

\bea\label{bg} F_\beta' F_\gamma & = & X_{23}X_{13}X_{14}X_{12}X_{13}X_{43} \nn \\ 
& \times & nT^{13}_{1,3, v^{''}} nT^{12}_{1,2, v^{''}} nT^{12, 13}_{2, 1, 4} nT^{13, 34}_{2, 3, 4} nT^{13}_{v^{'''}, 1, 3} nT^{34}_{v^{'''}, 4, 3} \nn \\
& \times & nT_{0, 2, 3} nT_{0, 1, 3} nT_{2, 3, 4} nT_{2, 1, 4} nT_{1, 3, v'} nT_{1, 4, v'}. \eea

The notation requires explanation. The vertices are numbered as shown in fig(\ref{braiding}). With this numbering $X_{12}$, for example, denotes the $X$ Pauli matrix acting on the link between vertices numbered 1 and 2. We also have 
\bea nT_{0, 2, 3} nT_{0, 1, 3} & = &\left[ n_1 T_{02}^1T_{12}^1T_{23}^1 + n_2 T_{02}^1T_{12}^1T_{23}^{-1} + n_3 T_{02}^1T_{12}^{-1}T_{23}^1 + n_4 T_{02}^1T_{12}^{-1}T_{23}^{-1} \right.\nn \\ 
& + & \left. n_5 T_{02}^{-1}T_{12}^1T_{23}^1 + n_6 T_{02}^{-1}T_{12}^1T_{23}^{-1} + n_7 T_{02}^{-1}T_{12}^{-1}T_{23}^1 + n_8 T_{02}^{-1}T_{12}^{-1}T_{23}^{-1} \right] \nn \\
& \times & \left[ n_1 T_{01}^1T_{12}^1T_{13}^1 + n_2 T_{01}^1T_{12}^1T_{13}^{-1} + n_3 T_{01}^1T_{12}^{-1}T_{13}^1 + n_4 T_{01}^1T_{12}^{-1}T_{13}^{-1} \right. \nn \\ 
& + & \left. n_5 T_{01}^{-1}T_{12}^1T_{13}^1 + n_6 T_{01}^{-1}T_{12}^1T_{13}^{-1} + n_7 T_{01}^{-1}T_{12}^{-1}T_{13}^1 + n_8 T_{01}^{-1}T_{12}^{-1}T_{13}^{-1}\right]. \eea
 
And we have
\bea nT^{23}_{0,2,3} nT^{13}_{0,1,3} & = & \left[ n_1 T_{02}^1T_{12}^1T_{23}^{-1} + n_2 T_{02}^1T_{12}^1T_{23}^1 + n_3 T_{02}^1T_{12}^{-1}T_{23}^{-1} + n_4 T_{02}^1T_{12}^{-1}T_{23}^1 \right.\nn \\ 
& + & \left. n_5 T_{02}^{-1}T_{12}^1T_{23}^{-1} + n_6 T_{02}^{-1}T_{12}^1T_{23}^1 + n_7 T_{02}^{-1}T_{12}^{-1}T_{23}^{-1} + n_8 T_{02}^{-1}T_{12}^{-1}T_{23}^1 \right] \nn \\
& \times & \left[ n_1 T_{01}^1T_{12}^1T_{13}^{-1} + n_2 T_{01}^1T_{12}^1T_{13}^1 + n_3 T_{01}^1T_{12}^{-1}T_{13}^{-1} + n_4 T_{01}^1T_{12}^{-1}T_{13}^1 \right. \nn \\ 
& + & \left. n_5 T_{01}^{-1}T_{12}^1T_{13}^{-1} + n_6 T_{01}^{-1}T_{12}^1T_{13}^1 + n_7 T_{01}^{-1}T_{12}^{-1}T_{13}^{-1} + n_8 T_{01}^{-1}T_{12}^{-1}T_{13}^1\right]. \eea

It is easy to see from the expression for the operator creating the flux in Eq.(\ref{f1}) that the two equations Eq.(\ref{gb}) and Eq.(\ref{bg}) are the same. Hence the flux excitation is bosonic. 

For the case when $n_1=-n_2$ we have the following operator that creates the flux excitation.
\beq \label{f2} F_{l_1} = n_1^2 X_{l_1}\left[\left(\frac{1+k^2}{2}\right) + \left(\frac{1-k^2}{2}\right)Z_{l_1}\right]Z_{l_2}Z_{l_3}Z_{l_2'}Z_{l_3'}.\eeq

The fusion rule for this operator is the same as the fusion rule of the operator in Eq.(\ref{f1}). It is easy to check using the expression for the flux operator in Eq.(\ref{f2}) in Eq.(\ref{bg}) and Eq.(\ref{gb}) that the flux operator is bosonic in this case as well. 

In order to obtain the double semion phase we need to use the following form for the modified vertex operator
\beq\label{asem} A_v = \sum_{l_1, l_2, l_3\in \{1, -1\}} X_{l_1}X_{l_2}X_{l_3} \alpha\left(l_2^{-1}l_1^{-1}, -l_2, -1\right)\alpha\left(-l_2, -1, l_3\right)\alpha\left(-l_1, -1, l_3\right)^{-1}T_{l_1}^{l_1} T_{l_2}^{l_2} T_{l_3}^{l_3}\eeq
where $\alpha\left(l_1, l_2, l_3\right)$ is an element of the cohomology group $H^3\left(\mathbb{Z}_2, U(1)\right)$. The cohomology group $H^3\left(\mathbb{Z}_2, U(1)\right)$ is given by $\mathbb{Z}_2$. Hence there are two cocycles one of which is the trivial cocycle. It is easy to see that for the trivial cocycle the operator $A_v$ in Eq.(\ref{asem}) reduces to the operator of the toric code. For the non-trivial cocycle $\alpha\left(-1, -1, -1\right)= -1$, the rest are equal to the identity element. The operator in Eq.(\ref{asem}) is precisely the same as the one used to define the twisted quantum double model in~\cite{YSWu} for the $\mathbb{C}(\mathbb{Z}_2)$ case. 

The two-link operators that are used to obtain the modified vertex operator in Eq.(\ref{asem}) are given by 
$$ \sum_{l_1, l_2} \left[\alpha\left(l_2^{-1}l_1^{-1}, -l_2, -1\right) T_{l_1}^{l_1}T^{l_2}_{l_2}\right],$$
$$ \sum_{l_2, l_3} \left[\alpha\left(-l_2, -1, l_3\right) T_{l_2}^{l_2}T_{l_3}^{l_3}\right] $$
and 
$$ \sum_{l_3, l_1}\left[\alpha\left(-l_1, -1, l_3\right) T_{l_3}^{l_3}T_{l_1}^{l_1}\right].$$

The expressions for the modified flux operator on a single link $l_1$ is given by
\beq F_{l_1} = X_{l_1}\left[\sum_{l_1, l_2, l_3}a_{l_1l_2l_3}T^{l_1}_{l_1} T^{l_2}_{l_2} T^{l_3}_{l_3}\right] \left[\sum_{l_1, l_2', l_3'}b_{l_1l_2'l_3'}T^{l_1}_{l_1} T^{l_2'}_{l_2'} T^{l_3'}_{l_3'}\right]\eeq
where $a_{l_1l_2l_3}$ and $b_{l_1l_2'l_3'}$ are functions of the variables on the links $l_1$, $l_2$, $l_3$, $l_2'$ and $l_3'$. The action of this string is shown in fig(\ref{triangle}). Using the convention of~\cite{YSWu} we order the vertices as follows
$v<v_1<v_2<v_3$. The operator creating the flux, $F_{l_1}$ has to commute with $A_v$, $A_{v_1}$, $A_{v_2}$ and $A_{v_3}$. This imposes the following conditions on the coefficients $a_{l_1l_2l_3}$ and $b_{l_1l_2'l_3'}$
\bea  \frac{a_{-l_1-l_2-l_3}b_{-l_1l_2'l_3'}}{a_{l_1l_2l_3}b_{l_1l_2'l_3'}} & = & \frac{\alpha\left(-l_2^{-1}l_1^{-1}, l_1, -1\right)\alpha\left(l_1, -1, l_3\right)}{\alpha\left(l_2^{-1}l_1^{-1}, -l_1, -1\right)\alpha\left(-l_1, -1, l_3\right)}  \\ 
\frac{a_{-l_1l_2l_3}b_{-l_1-l_2'-l_3'}}{a_{l_1l_2l_3}b_{l_1l_2'l_3'}} & = & \frac{\alpha\left(-l_2', -1, -l_1\right)\alpha\left(-1, -l_1, -l_1^{-1}(l_3')^{-1}\right)}{\alpha\left(-l_2', -1, l_1\right)\alpha\left(-1, l_1, l_1^{-1}(l_3')^{-1}\right)} \\
\frac{a_{l_1-l_2l_3}b_{l_1-l_2'l_3'}}{a_{l_1l_2l_3}b_{l_1l_2'l_3'}} & = & 1. \eea
 
 Using these conditions it is easy to check that the fusion and braiding rules for the operator creating the fluxes is that of the double semion phase.

\section{Remarks}

We have used modified transfer matrices of 3D lattice gauge theories with discrete gauge groups to find exactly solvable quantum models in two dimensions. The Abelian cases were studied with the example of the $\mathbb{Z}_n$ groups and were shown to generate a quantum phase identical to the quantum double phase based on the group algebra of these Abelian groups. This was seen by studying the excitations of these systems. The excitations were the same as the $\mathbb{Z}_n$ anyon models. The only change was in the operator creating these excitations. In the case when the vertex operators were modified the flux excitations were created by new operators whereas in the case where the plaquette operators were modified the operators creating the charge excitations were modified. Thus these models present another example of a system hosting $\mathbb{Z}_n$ anyons on the two dimensional lattice. There have been other examples of exactly solvable lattice models which host the $\mathbb{Z}_n$ anyons~\cite{exact}. The important point about the models presented in this paper is that they involve terms which are usually added as perturbations on the links to induce quantum phase transitions of the $\mathbb{Z}_n$ quantum double models. The models remain solvable in spite of including these terms. The only difference is that the modified vertex and plaquette operators obey a deformed version of the quantum double algebra. It is interesting question to find out if a suitable choice of a representation category can generate these models using the string-net formalism~\cite{stringLevin}.

The story changes when the discrete groups are non-Abelian. The example of the dihedral group $S_3$ was studied and the phase was found to be different from that of the corresponding quantum double phase. In particular the phase was found to be a modified version of the $S_3$ quantum double phase. The Abelian charge was no longer an excitation of the system instead it condensed to the ground state. The flux excitations corresponding to the $[\tau]$ conjugacy class is not existent that is cannot be expanded in the set of excitations of the system. Similar models were studied in~\cite{Del1} where they add the magnetic terms to the links just like perturbations. The plaquette operators were considered to be projectors to a normal subgroup of the full gauge group $G$. The resulting phases in~\cite{Del1} are condensed versions of the $S_3$ quantum double phase.  

The construction of the transfer matrices for two dimensional lattice gauge theories was shown in~\cite{QDMST}. In particular it was shown to contain the quantum double models for any involutory Hopf algebra. The models in~\cite{QDMST} were found by trivializing the $z_T$ and $z_S^*$ parameters. This implies there were no perturbation terms included in the transfer matrices considered in~\cite{QDMST}. The models considered in this paper can be considered as an extension of the formalism presented in~\cite{QDMST} where we make non-trivial choices for the parameters $z_T$ and $z_S^*$. Condensed phases of Abelian quantum double models were exhibited in~\cite{QDMST}. The models considered in this paper can be regarded as another way to obtain condensed phases along lines similar to~\cite{Del1}. 

The models considered in this paper included only four spin couplings. We could include more couplings which will result in more complicated models. For example if we couple a plaquette with the legs of the plaquette we can obtain models similar in spirit to the double semion model~\cite{SimonSemion}. However we cannot use the transfer matrix in Eq.(\ref{tz2}) to obtain the double semion model. The reason being that there are no choices of parameters $z_S$, $z_T$ which will result in this model. This calls for the construction of more general transfer matrices which can lead to these kinds of models. In section 5 we showed how we can go about finding the double semion phase by including two-qudit operators in the transfer matrix of lattice the gauge theory. We only worked out the case of $\mathbb{C}(\mathbb{Z}_2)$ in this section. For other groups this reduces to the twisted quantum double models constructed in~\cite{YSWu}. These models can be thought of as arising due to the introduction of new parameters to the vertices which we denoted as $z_{\textrm{ver}}^*$. Hence it is natural to use $z_{\textrm{vol}}$ to obtain the models dual to the twisted quantum double models. We will explore these in future works. 

It is interesting to note that the transfer matrix approach provides a variety of lattice gauge models including topological and non-topological ones. This approach may very well be appropriate for studying quantum phase transitions in these lattice models.  

\section*{Acknowledgements}
PP, MJBF and PTS thank FAPESP for support of this work. JPIJ thanks CNPq for support during this work.

\end{document}